\documentclass[aps,prl,twocolumn,superscriptaddress]{revtex4}

\usepackage{braket}
\usepackage{amsmath}
\usepackage{amssymb}

\newcommand{\PPS}[2]{{}_{t_2}\!\!\bra{#1}\otimes\ket{#2}\!{}_{t_1}}
\newcommand{\PPSd}[2]{{}_{{t_1}\!\!{}^\dagger}\!\bra{#2}\otimes\ket{#1}\!{}_{{t_2}\!{}^\dagger}}
\newcommand{\contract}[3]{\underline{#1}^{#2}\bullet\underline{#3}}
\newcommand{\contractb}[4]{\underline{#1}^{#2}\bullet\underline{#3}^{#4}}
\newcommand{\contractd}[3]{(\underline{#1}^{#2}\otimes\underline{#1}^{#2\dagger})\bullet\underline{#3}}
\newcommand{\contracte}[4]{\underline{#1}^{#2}\bullet\underline{#3}^{#4}}
\newcommand{\contractf}[4]{(\underline{#1}^{#2}\otimes\underline{#1}^{#2\dagger}) \bullet(\underline{#3}^{#4}\otimes\underline{#3}^{#4\dagger})}
\newcommand{\DME}[2]{\PPS{#1}{#2}\!\!\otimes\!\PPSd{#1}{#2}}

\begin{document}

\title{Pre- and post-selected quantum states: density matrices, tomography, and Kraus operators}

\author{Ralph Silva}\affiliation{H.H. Wills Physics Laboratory, University of Bristol, Tyndall Avenue, Bristol, BS8 1TL, U.K.}
\author{Yelena Guryanova}\affiliation{H.H. Wills Physics Laboratory, University of Bristol, Tyndall Avenue, Bristol, BS8 1TL, U.K.}
\author{Nicolas Brunner}\affiliation{H.H. Wills Physics Laboratory, University of Bristol, Tyndall Avenue, Bristol, BS8 1TL, U.K.}\affiliation{D\'epartement de Physique Th\'eorique, Universit\'e de Gen\`eve, 1211 Gen\`eve, Switzerland}
\author{Noah Linden}\affiliation{School of Mathematics, University of Bristol, University Walk, Bristol BS8 1TW, U.K.}
\author{Anthony J. Short}\affiliation{H.H. Wills Physics Laboratory, University of Bristol, Tyndall Avenue, Bristol, BS8 1TL, U.K.}
\author{Sandu Popescu}\affiliation{H.H. Wills Physics Laboratory, University of Bristol, Tyndall Avenue, Bristol, BS8 1TL, U.K.}

\begin{abstract}

We present a general formalism for charecterizing 2-time quantum states, describing pre- and post-selected quantum systems. The most general 2-time state is characterized by a `density vector' that is independent of measurements performed between the preparation and post-selection. We provide a method for performing tomography of an unknown 2-time density vector. This procedure, which cannot be implemented by weak or projective measurements, brings new insight to the fundamental role played by Kraus operators in quantum measurements. Finally, after showing that general states and measurements are isomorphic, we show that any measurement on a 2-time state can be mapped to a measurement on a preselected bipartite state.

\end{abstract}

\maketitle

Post-selection of states has provided us with a novel outlook on quantum mechanics, both with the possibility that the universe itself has a final post-selection, and with a new description of the information accessible for a quantum state. Although the concept of post-selection was described as far back as 1964 by Aharanov, Bergmann and Lebowitz \cite{ABL}, the discovery of weak measurements \cite{AAV} provoked renewed interest in the field. When such `non-disturbing' measurements are performed on a pre- and post-selected system, astonishing effects occur. For instance, the expectation value of a weak measurement of the spin of a spin-half system may be as large as 100 for a judiciously chosen preparation and post-selection of the spin state\cite{AAV}.

While initially controversial, weak measurements on post-selected states have since been used as a powerful tool for exploring the foundations of quantum mechanics \cite{hardyweak,kocsis,lundeen}. The concept was also explored experimentally \cite{ritchie} and shown to be useful in a wide range of contexts, ranging from superluminal light propagations \cite{superluminal,solli} to cavity QED \cite{wiseman}. More recently, weak measurements on post-selected states have found surprising applications in metrology, with the development of novel amplification techniques for precision measurements \cite{hosten,dixon,BS,FXS}.

In parallel, the ideas of pre- and post-selected systems were also developed from a conceptual point of view, leading to new ideas on the notion of time in quantum mechanics \cite{multi1}. Previous work has considered the case of pure pre- and post-selected states, both direct products as well as states entangled between the preparation and post-selection. A natural problem is to extend these to ensembles; this is the subject of the present paper. However, this is not equivalent to the case of generalizing preselected states to ensembles, since the success of post-selection affects the proportions of each state in the ensemble differently.

Here, we discuss the physical realization of a mixture of pre- and post-selected, or `2-time' states. We arrive at an equation that describes the probability statistics of any measurement made on this mixture between the preparation and post-selection. Using the formalism of 2 times, we then show that it is possible to describe such a mixture by a ``density vector'' that contains all of the information required to calculate measurement statistics. This density vector is independent of the choice of measurements made between the preparation and post-selection. We then provide a method for performing tomography on a 2-time mixture. This is shown to strictly require non-projective operators, unlike the case of performing tomography on preselected states.

Interestingly when we consider applying weak measurements to such mixtures, we find that they are insufficient to characterize the density vector, and in fact cannot distinguish between pure and mixed 2-time states. This is in stark contrast to the case of standard preselected states, where weak measurements are sufficient to perform full tomography.

Furthermore, we generalize the relation between preparations and measurements that was begun in \cite{APV}. Using the density vector that we derive and a suitably constructed ``Kraus density vector" that describes coarse-grained measurement outcomes, we prove a full isomorphism between preparations of 2-time ensembles and measurement outcomes, that also demonstrates the inherent 2-time nature of a general measurement.

Finally, we show that the 2-time density vector that we construct is equivalent to a bipartite density matrix for preselected states, and furthermore that any measurement on a 2-time state is isomorphic to a measurement on the corresponding bipartite state.

\emph{Creating a 2-time state.}---
The simplest manner of creating a pre- and post-selected state, as demonstrated in \cite{ABL}, is the following. The preparer of the state, henceforth referred to as Alice, prepares a state $\ket{\psi}$ at time $t_1$. She then passes the system to an `observer' who may perform any measurement he wishes to. The system is returned to Alice, who then performs a projective measurement with the state $\ket{\phi}$ as one of the outcomes. Only if this outcome is obtained, does the observer keep the results of his measurement.

It is straighforward to calculate that if the observer were to perform a measurement between the preparation and the post-selection described by the Kraus operators $\{A^\mu\}$   satisfying the completeness relation $\sum_\mu A^{\mu\dagger} A^\mu = I$ \footnote{For simplicity, we first consider  measurements in which each outcome corresponds to a single Kraus operator. Later, we consider the  general case in which a single outcome can be associated with multiple Kraus operators}, the probability of obtaining a particular outcome $\mu$ given that the post-selection succeeded would be\cite{APV}:
\begin{equation}\label{PRproduct}
	P(\mu|S,M) = \frac{|\bra{\phi}A^\mu\ket{\psi}|^2}{\sum_\nu|\bra{\phi}A^\nu\ket{\psi}|^2 }
\end{equation}
Throughout this paper $P(\mu|S,M)$ will denote the probability of the outcome $\mu$ given the success of the post-selection($S$), and that the measurement($M$) was performed between the preparation and post-selection.

An intuitive way to interpret the above expression is to consider the object $\PPS{\phi}{\psi}$ as a 2-time state, denoted as $\underline{\Psi}$. Here the forward evolving `ket' state $\ket{\psi}$ is a vector in the Hilbert space ${\cal{H}}_{t_1}^{\uparrow}$ and the backward evolving `bra' state $\bra{\phi}$ is a vector in the Hilbert space ${\cal{H}}_{t_2}^{\downarrow}$. The arrows denote the direction in time that the state is evolving in.

Thinking of $\PPS{\phi}{\psi}$ as a state in the joint Hilbert space ${\cal{H}}_{t_2}^{\downarrow}\otimes{\cal{H}}_{t_1}^{\uparrow}$ has the advantage of immediately suggesting the generalization to superpositions of pre- and post-selections, such as $\alpha\;\PPS{\phi_1}{\psi_1} +\beta\;\PPS{\phi_2}{\psi_2}$, or more generally, $\sum_{ij}\alpha_{ij}\;\PPS{i}{j}$. Following from eq(\ref{PRproduct}), if the observer performs a measurement on this state (between $t_1$ and $t_2$), we would expect the probabilities to take the form:
\begin{equation}\label{PRentangled}
	P(\mu|S,M) = \frac{|\sum_{ij}\alpha_{ij}\bra{i}A^\mu\ket{j}|^2}{\sum_\nu|\sum_{mn}\alpha_{mn}\bra{m}A^\nu\ket{n}|^2}
\end{equation}

In fact such superpositions can be physically realized\cite{ABL}, by using an ancilla to entangle the preparation and post-selection. Consider that Alice prepares the system and an ancilla in the superposition $\ket{\Psi}_{SA} = \sum_{ij}\alpha_{ij}\ket{j}_S\ket{i}_A$ at time $t_1$. She keeps the ancilla without disturbing it, and then at the later time $t_2$, she post-selects on the maximally entangled state $\ket{\Psi^+}_{SA}=(\sqrt{d})^{-1}\sum_k\ket{k}_S\ket{k}_A$ (here $d$ is the dimension of the system Hilbert space). This results in the 2-time state $\sum_{ij}\alpha_{ij}\;\PPS{i}{j}$ , that obeys eq(\ref{PRentangled}) for the statistics of any measurement made by the observer. The post-selection on a Bell state is an example of entanglement swapping, in this case from entanglement between the system and ancilla for entanglement between the system and itself at a later time.

In fact we may consider the Kraus operators $\{A^\mu\}$ themselves to be vectors in a product Hilbert space, that involve both forward and backward evolving states:
\begin{equation}\label{Krauss}
	\underline{A}^\mu \in {\cal{H}}_{t_1}^{\downarrow}\otimes {\cal{H}}_{t_2}^{\uparrow}\;,\;\;\;\underline{A}^\mu = \sum_{ij} A^\mu_{ij}\ket{i}\!{}_{t_2}\otimes{}_{t_1}\!\!\bra{j}
\end{equation}

This captures the `measure and prepare' nature of the Kraus operator, i.e. the ${}_{t_1}\!\!\bra{j}$ measures a forward evolving state at $t_1$ and a forward evolving state $\ket{i}_{t_2}$ is prepared at a later time $t_2$. If we now define the operation $\contract{A}{\mu}{\Psi}$ between 2-time vectors to be the tensor product $\underline{A}^\mu\otimes\underline{\Psi}$, followed by the contraction of any bra and ket of the same time, then given the Kraus operator $A^\mu=\sum_{ij}A^\mu_{ij}\ket{i}\bra{j}$, and the 2-time state $\underline{\Psi}=\sum_{ij}\alpha_{ij}\;\PPS{i}{j}$, we have that $\contract{A}{\mu}{\Psi} = \sum_{ij}\alpha_{ij}\bra{i}A^\mu\ket{j}$, and thus the probability expression in eq(\ref{PRentangled}) takes on the simple form:
\begin{equation}\label{PRnew}
	P(\mu|S,M) = \frac{\big|\contract{A}{\mu}{\Psi}\big|^2}{\sum_\nu \big|\contract{A}{\nu}{\Psi}\big|^2}
\end{equation}
Indeed, from the operation $(\bullet)$, we see immediately that the space of Kraus 2-time vectors is dual to the space of pure 2-time states.

\emph{Mixing pre- and post-selected states.}---
Thus, using the formalism of 2-time states, superpositions of pre- and post-selections have been understood successfully. The open question at this point is, can we do the same for mixtures or ensembles of 2-time states?

Consider the ensemble described by the set $\{p^r, \underline{\Psi}^r\}$, where the $p^r$ are normalized ($\sum_r p^r=1$). We interpret the ensemble as the following. Alice has a number of pre- and post-selection procedures indexed by $r$, where the $r^{th}$ procedure results in the pure 2-time state $\underline{\Psi}^r$. She performs the $r^{th}$ procedure with the probability $p^r$.

What are the statistics of such an ensemble for the outcomes of a measurement performed by an observer between the preparation and post-selection? For preselected states, the probability for an ensemble is simply the weighted average of the probabilities for pure states, however this is \emph{not} the case for 2-time states (The weighted average of eq \ref{PRnew} is incorrect). This is because the post-selection affects the proportions of each pure 2-time state in the ensemble differently. Taking this into account, the statistics are shown to obey (Appendix A) the following expression:
\begin{equation}\label{PR}
	P(\mu|S,M)	=  \frac{\sum_r p^r \big|\contracte{A}{\mu}{\Psi}{r}\big|^2}{\sum_\nu\sum_s p^s\big|\contracte{A}{\nu}{\Psi}{s}\big|^2}
\end{equation}

Note that in the above ensemble, the probabilities $p^r$ refer to the relative frequencies of \emph{attempting} the procedure for each 2-time state. Of course each 2-time state has a different chance of being successful at the post-selection stage, and this depends in general upon the measurement that the observer chooses to perform.

We may imagine a different interpretation of the ensemble, where the probabilities refer to the proportions of \emph{successfully} post-selected 2-time states present in the ensemble. However this requires a post-selection procedure that is dependent upon the measurement chosen by the observer, and hence is not a natural definition of a 2-time ensemble. (For more details see Appendix B)

\emph{Density Matrix for 2-time Ensembles.}---
We've demonstrated a physical realization for an ensemble of 2-time states. However, we know that for standard preselected states, knowing the entire preparation ensemble is unnecessary to calculate measurement outcomes and probabilities, rather there exists a density matrix that contains all of the required information. In fact, the density matrix is more useful than the original ensemble since many ensembles can give rise to the same density matrix. Does there exist a similar object for 2-time states?

To proceed, we need to introduce some notation for clarity. Recall for standard preselected states that from the pure state $\ket{\psi}$ we obtain a density matrix $\ket{\psi}\bra{\psi}$. Here the bra $\bra{\psi}$ does not represent a backward evolving state, but rather the dual of a forward evolving state at the same time. We will therefore label it as ${}_{t^\dagger}\!\!\bra{\psi}$, to differentiate it from a backward evolving state, and thus the density matrix for a pure preselected state at time $t$ will be $\ket{\psi}_{tt^\dagger}\!\!\bra{\psi}$.

In the 2-time formalism, we define the ``density \emph{vector}'' instead, analogous to the above. For a pure 2-time state $\PPS{\phi}{\psi}$ the density vector is given by $\DME{\phi}{\psi}$, and in general, for the 2-time state $\underline{\Psi}$, the density vector is $\underline{\Psi}\otimes\underline{\Psi}^\dagger$. Given an ensemble of 2-time states $\{ p^r,\underline{\Psi}^r\}$, we expect the density vector to be:
\begin{equation}\label{densityvector}
	\underline{\eta} = \sum_{r} p^r(\underline{\Psi}^r\otimes\underline{\Psi}^{r\dagger})
\end{equation}

 The density vector is a state in the Hilbert space ${\cal{H}}_{t_2}^{\downarrow}\otimes{\cal{H}}_{t_1}^{\uparrow}\otimes{\cal{H}}_{t_1}^{\uparrow\dagger}\otimes {\cal{H}}_{t_2}^{\downarrow\dagger}$. From its construction, the density vector is ``positive'', in the sense that\footnote{We later prove that $\eta$ is isomorphic to a bipartite operator that is positive in the usual sense}
\begin{equation}\label{positivity}
\contractd{V}{}{\eta}\geq 0 \;\;\;\forall \;\;\;\underline{V}\in {\cal{H}}_{t_2}^{\uparrow}\otimes{\cal{H}}_{t_1}^{\downarrow}
\end{equation}
The operation $\bullet$ is defined as before, with the rule that dagger bras only contract dagger kets of the same time.

Given the density vector $\underline{\eta}$ that describes an ensemble of 2-time states, we would expect the statistics of a measurement $\{A^\mu\}$ between $t_1$ and $t_2$ to obey:
\begin{equation}\label{PRmatrix}
	P(\mu|S,M) =  \frac{\contractd{A}{\mu}{\eta}}{\sum_\nu\contractd{A}{\nu}{\eta}}
\end{equation}

Expanding the above, and noting that $\big|\contractb{A}{\mu}{\Psi}{r}\big|^2 = \contractf{A}{\mu}{\Psi}{r}$, we arrive at the correct expression for the probability already derived in eq (\ref{PR}). Note that only $\underline{\eta}$ and not the detailed construction of the ensemble appears in the formula for the probability. Thus the density vector performs the same role for ensembles of 2-time states as the density matrix does for standard preselected ensembles.

In addition, although we provided a particular construction of a 2-time ensemble, any different preparation and post-selection procedure would also be described by a density vector of the form above. The construction provided is only useful in so far as it can be used to create any 2-time ensemble.

\emph{Tomography of the Density Vector.}---
Now that we can describe any 2-time ensemble by a density vector, we consider the problem of determining the density vector of an unknown ensemble, i.e. performing tomography.

To determine the density matrix of standard preselected states, projective measurements are sufficient for performing tomography. Interestingly, this is not the case for 2-time states. Indeed, consider the pure qubit 2-time superposition $1/\sqrt{2} \;(\PPS{0}{0}+\PPS{1}{1})$, where $\ket{0}$ and $\ket{1}$ are orthogonal spin states along a chosen axis. If we consider a projector along any direction on the Bloch sphere, $\hat{P}=\ket{\theta,\phi}\bra{\theta,\phi}$, the relative probability for the outcome corresponding to this projector will be independent of the direction. From (eq \ref{PRentangled}), the numerator of the expression for the probability of the outcome corresponding to $\ket{\theta,\phi}$ will be
\begin{eqnarray}
	N &=& |\braket{0|\theta,\phi}\braket{\theta,\phi|0} + \braket{1|\theta,\phi}\braket{\theta,\phi|1}|^2 \nonumber\\
		&=& (|\braket{0|\theta,\phi}|^2 + |\braket{1|\theta,\phi}|^2)^2 = 1
\end{eqnarray}
Since the denominator is the sum of two such terms, we see that the probabilities are equal to $1/2$ for all projectors. Consider as well the equal mixture of the 2-time states $\PPS{0}{0}$ and $\PPS{1}{1}$. By a similar calculation (eq \ref{PR}), we see that the probability of an outcome corresponding to any projector is $1/2$, and thus projective measurements cannot distinguish this mixture from the pure state above. Thus we conclude that measurements involving non-projective Kraus operators are vital for distinguishing 2-time ensembles, which is further indication of the 2-time nature of Kraus operators. In the example above, the Kraus operator $\ket{-}\bra{+}$ returns zero probability for the equal superposition of  $\PPS{0}{0}$ and $\PPS{1}{1}$ but non-zero for the equal  mixture.

In order to perform tomography on a general 2-time ensemble, we simply pick a set of Kraus operators that provide a number of linear combinations of the elements of the density vector that are sufficient to reconstruct it. An explicit construction is as follows: for a system of dimension $d$, with an orthonormal basis of states $\ket{i}$, consider the $d^2$ unique operators $O_{ij}=\ket{i}\bra{j}$. The measurement described by the normalized set of $4d^4$ Kraus operators $ \frac{1}{\sqrt{8d^3}}\{O_{ij}\pm O_{kl}, O_{ij}\pm iO_{kl}\}$ is sufficient to determine every element of the 2-time density vector.

Interestingly, like projective measurements, weak measurements are not sufficient to perform tomography on the most general 2-time ensemble. In fact, as shown in Appendix C, weak measurements are incapable of differentiating between pure 2-time states and ensembles of 2-time states.

\emph{Kraus density vectors.}---
In all of the probability calculations in earlier sections, we assumed that our measurement is a detailed one, i.e. we completely read the measuring device, and thus to each outcome $\mu$ of the measurement corresponds a single Kraus operator $A^\mu$. However we may also consider coarse grained measurements, which can be achieved either by a) making a detailed measurement but lumping outcomes together, or b) by not completely reading the measurement device, leaving the device still entangled to the system. This is reminiscent of the two ways of producing mixed states, by either forgetting which state we prepare out of a set, or by preparing particles entangled with ancillae.

For a coarse-grained measurement, to each outcome $\mu$ corresponds a set of Kraus operators $\{A^\mu_\chi\}$. The probability of the outcome $\mu$ is obtained by calculating the `virtual' probabilities obtained for the detailed measurement, and then summing those over the index $\chi$:
\begin{equation}\label{PRcourse}
	P(\mu|S,M) = \sum_\chi P(\mu,\chi|S,M)
\end{equation}

This presents the following interesting question. Consider two coarse-grained measurements, both having the same set of outcomes $\mu$, but described by different sets of Kraus operators $A^\mu_\chi$ and $B^\mu_\lambda$ respectively. When are these measurements identical in that they give the same probability for the every outcome $\mu$ when acting on a general 2-time ensemble? This is closely related to the question of when two quantum channels are identical, and indeed any channel  can be thought of as a coarse-grained measurement with a single outcome. To address these issues we define a ``Kraus density vector":
\begin{equation}\label{2-time POVM}
	\underline{K}^\mu	 = \sum_{\chi}	\underline{A}^\mu_\chi\otimes\underline{A}^{\mu\dagger}_\chi
\end{equation} with each Kraus 2-time vector $\underline{A}^\mu_\chi$ derived from the Kraus operator as $\underline{A}^\mu_\chi = \sum_{ij} A^\mu_{\chi,ij}\ket{i}\!{}_{t_2}\!\otimes{}_{t_1}\!\!\!\bra{j}$.

This is a vector in the product Hilbert space ${\cal{H}}_{t_2}^{\uparrow} \otimes {\cal{H}}_{t_1}^{\downarrow} \otimes {\cal{H}}_{t_1}^{\downarrow\dagger} \otimes {\cal{H}}_{t_2}^{\uparrow\dagger}$. The Kraus density vector is positive in a similar sense as the density vector of an ensemble of 2-time states (eq \ref{positivity}).

As we expect, two coarse-grained measurements are identical if and only if they have the same Kraus density vector for all of their outcomes. This can be seen by using the Kraus density vector to generalize the measurement statistics given in eq(\ref{PRmatrix}) to the case of coarse-grained measurements. Using eq(\ref{2-time POVM}), we have:
\begin{equation}\label{PRneat}
	P(\mu|S,M) = \frac{\contract{K}{\mu}{\eta}}{\sum_\nu \contract{K}{\nu}{\eta}}
\end{equation}

Identical to the duality of Kraus 2-time vectors $\underline{A}^\mu$ and pure 2-time states $\underline{\Psi}$, we see from the above that the space of Kraus density vectors is dual to the space of density vectors of 2-time ensembles.

\emph{Preparation-Measurement Equivalence.}--- From the view of the experimentalist, the preparation of a state involves exactly the same actions as a measurement. In both cases we act on the system with various devices and record the results of the measuring device. The difference is that when preparing a state, we consider only one of the outcomes of the measurement to be useful (the outcome corresponding to the state we wish to prepare). Thus a state and the measurement operator corresponding to a single outcome are equivalent.

However, in the standard formulation of quantum mechanics states and measurement outcomes are completely different animals; states are vectors in the Hilbert space while measurement outcomes are represented by Kraus operators acting on states. Thus there is a discrepancy between the formalism and the actions in the laboratory.

The above problem was noted in \cite{APV} and it was shown that the 2-time formalism puts states and measurements on an equal footing. A 2-time state expressed as $\sum_{ij}\alpha_{ij}\,\PPS{i}{j}$ is equivalent to the Kraus operator corresponding to a single outcome $\mu$ of a detailed measurement, $\sum_{ij}A^\mu_{ij}\ket{i}_{t_2 t_1}\!\!\!\bra{j}$, as is also seen through the duality between Kraus 2-time vectors and pure 2-time states, noted in an earlier section. However, this equivalence only covers \emph{pure} 2-time states and Kraus operators for \emph{detailed} measurements.

Once we generalize the description of 2-time states to include ensembles, and measurements to include coarse-grained outcomes, we see that the duality is maintained, between the 2-time density vector for an ensemble $\underline{\eta}$ and the Kraus density vector for a single outcome $\underline{K}^\mu$, and thus the description of the most general 2-time state is seen to be equivalent to the description of an outcome of the most general measurement.

\emph{Equivalence between 2-time preparations and measurements and the bipartite scenario.}--- Considering that 2-time states reside in a product Hilbert space naturally suggests a relationship with bipartite states, and this can seen through the following explicit isomorphism. Given a pure 2-time state expanded in a given basis $\underline{\Psi} = \sum_{ij}\alpha_{ij}\;\PPS{i}{j}$, we construct the bipartite state isomorphic to $\underline{\Psi}$ as $\ket{\Psi}_{AB} = \sum_{ij}\alpha_{ij}\ket{i}_A\otimes\ket{j}_B$.

Thus given an ensemble of 2-time states $\{p^r, \underline{\Psi}^r\}$, we can construct a bipartite density matrix $\rho_{AB} = \sum_r p^r \ket{\Psi^r}\bra{\Psi^r}$, that is isomorphic to the density vector for the ensemble $\underline{\eta} = \sum_r p^r (\underline{\Psi}^r\otimes \underline{\Psi}^{r\dagger})$. This establishes an isomorphism between 2-time and bipartite states.

This can be extended to the case of measurements as well. Given a Kraus 2-time vector in the same basis as before\footnote{The isomorphisms between 2-time and bipartite states and measurements are basis dependent}, $\underline{A}^\mu = \sum_{ij}A^\mu_{ij}\ket{i}\!{}_{t_2}\!\otimes\!{}_{t_1}\!\!\bra{j}$, we construct the bipartite vector $\ket{a^\mu}_{AB} = \sum_{ij} A^{\mu*}_{ij}\ket{i}_A\otimes\ket{j}_B$. If we consider a coarse-grained measurement outcome instead, then from the Kraus density vector $\underline{K}^\mu = \sum_{\chi} \underline{A}^{\mu}_{\chi}\otimes\underline{A}^{\mu\dagger}_{\chi}$, we have the isomorphic (and positive) bipartite operator $\tilde{E}^\mu_{AB} = \sum_{\chi} \ket{a^\mu_\chi}\bra{a^\mu_\chi}$.

The isomorphisms above have been constructed in such a way that given a Kraus density vector $\underline{K}^\mu$ and the density vector for an ensemble $\underline{\eta}$, and the corresponding isomorphic bipartite operator $\tilde{E}^\mu_{AB}$ and  density matrix $\rho_{AB}$, we have the equality $\contract{K}{\mu}{\eta} = tr(\tilde{E}^\mu\rho)$.

From the isomorphism, if we have a set of Kraus density vectors $\{\underline{K}^\mu\}$ that describes a measurement, we can construct an isomorphic set of positive bipartite operators. However, these do not form a complete measurement, since $\sum_\mu \tilde{E}^\mu \neq I_{AB}$. Conversely, given a normalized set of positive bipartite operators, the set of Kraus density vectors that we obtain from theabove isomorphism does not form a normalized measurement (see Appendix D for more details).

However, given a set of (non-complete) positive operators $\{\tilde{E}^\mu\}$, we may still ask the question: What are the relative probabilites associated with each operator, when measured on a given state? To answer this question with a measurement, we first complete the set by subnormalizing and adding an element -  $\sum_\mu c\tilde{E}^\mu + E^\prime = I$, where $E^\prime$ is a positive operator. We then perform the measurement corresponding to this set, and discard all of the outcomes corresponding to $E^\prime$. The resulting relative probability for the outcome $\mu$ from this measurement procedure on the state $\rho$ is given by:
\begin{equation}
	P(\mu) = \frac{tr(\tilde{E}^\mu\rho)}{\sum_\nu tr(\tilde{E}^\nu\rho)}
\end{equation}
The same procedure can be used to complete a set of Kraus density vectors that does not form a measurement on a 2-time state. In this case, the expression for the probability obeys eq(\ref{PRneat}), as before. But since $\contract{K}{\mu}{\eta} = tr(\tilde{E}^\mu\rho)$, this is identical to the above expression for the bipartite case.

Thus, by defining a general measurement to consist simply of a set of positive density vectors (or operators), we find that any scenario involving the creation and measurement of a 2-time state is equivalent to the preparation and measurement of a bipartite state.

\emph{Discussion}. --- We have presented a formalism for characterizing general 2-time states and measurements. These ideas are suitable for generalization to states and measurements on systems at multiple times, which involve both forward and backward evolving states \cite{APV}. For instance, consider the 4-time state $\underline{\Psi}={}_{t_4}\!\!\bra{\psi_4}\otimes{}_{t_3}\!\!\bra{\psi_3}\otimes\ket{\psi_2}\!{}_{t_2}\!\otimes\ket{\psi_1}\!{}_{t_1}$. The operation ($\bullet$) is also naturally generalized. For example, given a 2-time vector $\underline{\beta}=\ket{\beta_3}{}_{t_3} \!\!\otimes\! {}_{t_2}\!\!\bra{\beta_2}$, the operation $\contract{\beta}{}{\Psi}$ results in a 2-time vector.

\emph{Acknowledgements}. We thank M. Navascues for interesting discussions during this work. We also acknowledge financial support from the Swiss National Science Foundation (grant PP00P2 138917), the ERC (Advanced Grant “NLST”), and the Royal Society.

\section{APPENDIX}
\emph{A. Derivation of the probability of an outcome for a measurement on a 2-time ensemble between the preparation and post-selection.}---
Consider that we have an ensemble of 2-time states, $\{p^r,\underline{\Psi}^r\}$, with each 2-time state expanded in a given basis as $\underline{\Psi}^r = \sum_{ij}\alpha^r_{ij}\;\PPS{i}{j}$, and a detailed measurement that we perform between the preparation and post-selection, described by the complete set of Kraus operators $\{A_\mu\}$. We wish to calculate $P(\mu|S,M)$, the probability of the outcome $\mu$, given the success of the post-selection, and the measurement having been performed between $t_1$ and $t_2$. The notation $P(A,B|C,D)$ refers to the probability of events $A$ and $B$ given that events $C$ and $D$ are true.

The different events that we have to consider are the outcome ($\mu$), the success of the post-selection ($S$), the measurement having been performed ($M$), and the event that the 2-time system is in the single pure state $\underline{\Psi}^r$ out of the entire ensemble ($r$).

Using Bayesian statistics, we have
\begin{align}
	P(\mu|S,M) &= \sum_r  P(\mu|r,S,M)\cdot P(r|S,M)\label{B1}\\
	P(r|S,M) &= \frac{P(r,S|M)}{P(S|M)} = \frac{P(S|r,M) \cdot P(r|M)}{P(S|M)}\label{B2}
\end{align}
Considering that the measurement is described by the set of Kraus operators $\{A_\mu\}$, we can calculate the values of the above variables straightforwardly:
\begin{align}
	P(r|M) 	&= p^r\label{B3}\\
	P(S|r,M) 	&= \sum_\mu\Big|\sum_{ij}\alpha^r_{ij}\bra{i}A^\mu\ket{j}\Big|^2\\
			&= \sum_\mu \big|\contractb{A}{\mu}{\Psi}{r}\big|^2\label{B4}\\
	P(S|M)	&= \sum_\nu\sum_s p^s\Big|\sum_{ij}\alpha^s_{ij}\bra{i}A^\nu\ket{j}\Big|^2\\
			&= \sum_\nu\sum_s p^s\big|\contractb{A}{\mu}{\Psi}{r}\big|^2\label{B5}\\
	P(\mu|r,S,M)	&= \frac{\big|\contractb{A}{\mu}{\Psi}{r}\big|^2}{\sum_\nu \big|\contractb{A}{\nu}{\Psi}{r}\big|^2}\label{B6}
\end{align}

Combining the above expressions, we arrive at the final probability that we seek, $P(\mu|S,M)$, in eq (\ref{PR}).

\emph{B. Defining an ensemble of 2-time states.}---
When considering the definition of a 2-time ensemble, we wish to seperate the role of the `creator' of the ensemble, the one that prepares and post-selects, and the `observer' of the ensemble, the one that performs a measurement between the preparation and post-selection. This implies that our definition of the ensemble must be independent of the measurement performed by the observer.

In the laboratory, there are 2 naturally contrasting methods that suggest themselves physical realization of the ensemble of 2-time states $\{p^r, \underline{\Psi}^r\}$.

The method followed in the paper involves interpreting the $p^r$ as the proportions of the ensemble at the preparation stage. Thus $p^r$ is the relative frequency with which we attempt to post-select on state $\underline{\Psi}^r$. Of course, depending on the measurement performed by the observer the actual frequency will differ from the prepared frequency. The advantage of this manner of defining the ensemble is that it is clearly independent of what the observer chooses to measure, and significantly, one of the main results of the paper is to show that such an ensemble is described by a density vector, also independent of the measurement, with very useful consequences.

The other interpretation of the ensemble is the $p^r$ are the relative frequencies of the 2-time states after the post-selection. This can be enforced by retrospectively performing a statistical selection. For example, if we wish to mix different 2-time states in equal proportions and notice that some states succeed more than others, we may discard some of the successful post-selection outcomes in order to achieve an equal proportion. However, this procedure is dependent on the measurement made by the observer, and gives rise to difficulties, as demonstrated in the following example.

Consider the explicit case of preparing an ensemble of $\PPS{0}{0}$ and $\PPS{0}{1}$ mixed in equal proportion. If the observer conducts the measurement $M_1$ described by the (complete) Kraus operators $A^1=\ket{0}\bra{0}$ and $A^2=\ket{+}\bra{1}$, the first 2-time state always gives outcome 1, and the post-selection always succeeds ($|\bra{0}A^1\ket{0}|^2 = 1$), while the second 2-time state always gives outcome 2, but the post-selection after this outcome only succeeds half the time. ($|\bra{0}A^2\ket{1}|^2 = 1/2$). In this case the post-selector can simply discard half of the successful attempts at preparing the first 2-time state, in order to force the proportions to be equal.

However, consider instead that the observer chooses randomly between the above measurement and another measurement $M_2$ described by the complete Kraus operators $B^1=\ket{+}\bra{0}$ and $B^2=\ket{0}\bra{1}$. Once again, the first 2-time state only gives outcome 1, and the second only gives outcome 2, however in this case the first 2-time state only succeeds in being post-selected half the time, while the second always succeeds.

Thus if among an ensemble, the observer randomly chooses between these 2 measurements, the post-selector will not know which successful states to discard. In fact if the observer chooses $M_1$ 50\% of the time (and also $M_2$), each 2-time state will succeed equally overall. However the proportions corresponding to when $M_1$ was performed will be different to those when $M_2$ was performed.

Thus we see that interpreting the proportions of the 2-time states present in an ensemble as the proportions present after post-selection fails to provide a definition independent of the measurement the observer chooses to make between the preparation and post-selection.

\emph{C. Insufficiency of Weak measurements for Tomography of 2-time ensembles.}---
In the case of preselected states of finite dimensions, we can use weak measurements to perform a complete tomography on any ensemble. The weak value of an observable on a preselected state is simply the expectation value of the operator on the state. By using the appropriate observables, we can determine the density matrix of any preselected state.

Interestingly, in the case of 2-time states, weak measurements are not sufficient to perform tomography on the most general density vector. In fact weak measurements are incapable of differentiating between pure and mixed 2-time states, as we demonstrate here.

We recall from \cite{APV} the weak value of an observable $\hat{A}$ on a pure 2-time state $\underline{\Psi} = \sum_{ij}\alpha_{ij}\;\PPS{i}{j}$:
\begin{equation}\label{weakpure}
	A_w 	= \frac{\sum_{ij}\alpha_{ij}\bra{i}\hat{A}\ket{j}}{\sum_{mn}\alpha_{mn}\braket{m|n}}
\end{equation}

In fact if we convert the observable $\hat{A} = \sum_{ij}A_{ij}\ket{i}\bra{j}$ into a 2-time vector $\underline{A}=\sum_{ij}A_{ij}\ket{i}\!{}_{t_2}\!\otimes\!{}_{t_1}\!\!\bra{j}$, the weak value can be written in the simple form:
\begin{equation}\label{weakpurevector}
	A_w = \frac{\contract{A}{}{\Psi}}{\contract{I}{}{\Psi}}
\end{equation}
where the 2-time vector $\underline{I} = \sum_i \ket{i}\!{}_{t_2}\!\otimes\!{}_{t_1}\!\!\bra{i}$. Given this expression, we can determine any pure state $\underline{\Psi}$ by weak measurements on an appropriate set of observables, and thus weak measurements can distinguish perfectly among pure 2-time states.

If instead we have an \emph{ensemble} of 2-time states, $\{ p^r\!,\underline{\Psi}^r\}$, we can calculate the weak value of $\hat{A}$ as follows:
\begin{equation}\label{W1}
	A_w = \sum_r A_w^r \cdot P(r|S)
\end{equation}
Here $P(r|S)$ is the probability of the 2-time state being $\underline{\Psi}^r$ given the success of the post-selection, and  $A^r_w$ is the weak value of the observable $\hat{A}$ for that state. Using these, we can calculate the weak value:
\begin{align}	
	P(r|S)		&= \frac{p_r \big|\sum_{ij}\alpha^r_{ij}\braket{i|j}\big|^2}{\sum_s p_s \big|\sum_{mn}\alpha^s_{mn}\braket{m|n}\big|^2}\\
			&=  \frac{p^r\big|\contractb{I}{}{\Psi}{r}\big|^2}{\sum_s p^s\big|\contractb{I}{}{\Psi}{s}\big|^2}\label{W2}
\end{align}
\begin{equation}\label{W3}
	\therefore A_w = \frac{\sum_r p_r \big(\contractb{A}{}{\Psi}{r}\big)\big(\contractb{I}{\dagger}{\Psi}{r\dagger}\big)}{\sum_s p_s \big(\contractb{I}{}{\Psi}{s}\big)\big(\contractb{I}{\dagger}{\Psi}{s\dagger}\big)}
\end{equation}

Given the density vector $\underline{\eta}$ that describes the ensemble, we can simplify the above by considering the ``weak value vector" $\underline{\eta}_w$ of the ensemble to be the contraction
\begin{equation}
	\underline{\eta}_w = \contract{I}{\dagger}{\eta} = \sum_r p^r \underline{\Psi}^r\big(\contractb{I}{\dagger}{\Psi}{r\dagger}\big)
\end{equation}
Unlike all of the contractions considered previously, this is not a number, but rather a 2-time vector, since $\underline{I}^\dagger$ only contracts 2 of the 4 Hilbert spaces present in $\underline{\eta}$.

Using the weak value vector, the weak value of an observable $\hat{A}$ on a 2-time ensemble can be expressed simply as:
\begin{equation}
	A_w = \frac{\contract{A}{}{\eta}_w}{\contract{I}{}{\eta}_w}
\end{equation}

The immediate observation is that this is identical to form of eq(\ref{weakpurevector}). Indeed given any 2-time density vector $\underline{\eta}$, we can find a pure 2-time state that returns the same weak values. Explicitly we pick the unique 2-time state $\underline{\Psi} \propto \underline{\eta}_w$. Thus we conclude that weak measurements cannot distinguish between pure and mixed 2-time states.

As a sidenote, we observe that the density vector is sufficient to calculate the result of weak measurements as well.

\emph{D. Relation between 2-time and bipartite measurements}. --- Given a general coarse-grained measurement on a 2-time state, we derived the Kraus density vectors from the Kraus operators as $\underline{K}^\mu = \sum_{\chi} \underline{A}^\mu_\chi\otimes\underline{A}^{\mu\dagger}_\chi$. If the original set of Kraus operators is normalized ($\sum_{\mu,\chi} A^{\mu,\chi\dagger}A^{\mu,\chi} = I$), then there is an equivalent normalization condition for the Kraus density vectors:
\begin{equation}\label{normK}
	\sum_\mu \contract{K}{\mu}{I_2} = \underline{I_1}
\end{equation} where the 2-time vector $I_2 = \sum_i \ket{i}\!{}_{t_2^\dagger}\otimes{}_{t_2^{}}\!\!\bra{i}$, and $I_1 = \sum_i \ket{i}\!{}_{t_1^\dagger}\otimes{}_{t_1^{}}\!\!\bra{i}$.

How does this normalization reflect on the bipartite measurement isomorphic to the above 2-time measurement? Using the isomorphism, from the Kraus density vector $\underline{K}^\mu = \sum_{\chi} \underline{A}^{\mu}_{\chi}\otimes\underline{A}^{\mu\dagger}_{\chi}$ we obtained the positive bipartite operator $\tilde{E}^\mu_{AB} = \sum_{\chi} \ket{a^\mu_\chi}\bra{a^\mu_\chi}$. The normalization in eq (\ref{normK}) becomes, in the bipartite case:
\begin{equation}\label{normE}
	\sum_\mu tr_B(\tilde{E}^\mu_{AB}) = I_A
\end{equation}
This is not equivalent to the normalization of a set of POVM elements. Thus the set of positive operators that we obtain from the isomorphism has to be completed and normalized to form a normalized measurement.

Conversely, if we begin with a normalized set of POVM elements $\{E^\mu\}$, the set of Kraus density vectors $\{\underline{\tilde{K}}\}$ that we obtain from the isomorphism is exactly supernormalized, since the normalization condition for the bipartite case $\sum_\mu E^\mu_{AB} = I_{AB}$ translates to the 2-time case as:
\begin{equation}
	\sum_\mu \contract{\tilde{K}}{\mu}{I_2} = d\cdot\underline{I_1}
\end{equation}
Thus we can simply subnormalize the above set to obtain a valid 2-time measurement.

\end{document}